\documentclass[12pt,a4paper,dvips]{article}
\usepackage{epsfig,wrapfig,times,mathptm} 
\renewcommand{\section}[1]{\vspace{6pt} \noindent\mbox{#1} \newline \noindent}
\renewcommand{\subsection}[1]{\vspace{6pt} \noindent\mbox{\underline{#1}} 
\newline \noindent}
\renewcommand{\subsubsection}[1]{\vspace{6pt} \noindent\mbox{\underline{#1}}
\noindent}

\newfont{\sansb}{cmssbx10}
\newfont{\sans}{cmss10}

\begin{document}
{\small OG 4.2.6 \vspace{-24pt}\\}     
{\center \LARGE 
CONSTRAINTS ON COSMIC-RAY ORIGIN THEORIES FROM TEV GAMMA-RAY OBSERVATIONS
\vspace{6pt}\\}
R.W.Lessard$^1$, P.J.Boyle$^2$, S.M. Bradbury$^4$, J.H.Buckley$^3$, 
A.C.Burdett$^4$,J.Bussons Gordo$^2$,\\D.A.Carter-Lewis$^5$, M.Catanese$^5$,
M.F.Cawley$^6$, D.J.Fegan$^2$, J.P.Finley$^1$, J.A.Gaidos$^1$, A.M.Hillas$^4$,
\\F.Krennrich$^5$, R.C.Lamb$^7$, C.Masterson$^2$, J.E.McEnery$^2$, 
G.Mohanty$^5$, J.Quinn$^2$, A.J.Rodgers$^4$,\\H.J.Rose$^4$,F.W.Samuelson$^5$,
G.H.Sembroski$^1$, R.Srinivasan$^1$, T.C.Weekes$^3$ and J.Zweerink$^5$
\vspace{6pt}\\
{\it $^1$Department of Physics, 
         Purdue University, West Lafayette, IN 47907-1396, U.S.A.\\
     $^2$Department of Experimental Physics,
         University College Dublin, Belfield, Dublin 4, Ireland\\
     $^3$Whipple Observatory, Harvard-Smithsonian CfA, P.O. Box 97, 
         Amado, AZ 85645-0097, U.S.A., \\
     $^4$Department of Physics,
         University of Leeds, Leeds, LS2 9JT, Yorkshire, U.K., \\
     $^5$Department of Physics and Astronomy,
         Iowa State University, Ames, IA 50011-3160 USA, \\
     $^6$Department of Physics,
         Saint Patrick's College, Maynooth, Co. Kildare, Ireland,\\
     $^7$Space Radiation Laboratory, California Institute of Technology, 
         Pasadena, CA 91125 USA\\}
{\center ABSTRACT\\}
If supernova remnants (SNRs) are the sites of cosmic-ray acceleration,
the associated nuclear interactions should result in observable fluxes
of TeV gamma-rays from the nearest SNRs. Measurements of the gamma-ray
flux from six nearby, radio-bright, SNRs have been made with the
Whipple Observatory gamma-ray telescope.  No significant emission has
been detected and upper limits on the $>$300 GeV flux are reported.
Three of these SNRs (IC443, gamma-Cygni and W44) are spatially
coincident with low latitude unidentified sources detected with EGRET.
These upper limits weaken the case for the simplest models of shock
acceleration and energy dependent propagation.

\setlength{\parindent}{1cm}

\section{INTRODUCTION}
	It is generally believed that cosmic rays with energies less
than $\sim 100$ TeV originate in the galaxy and are accelerated in
shock waves in shell-type SNRs. This hypothesis is
supported by several strong arguments. First, supernova blast shocks are one
of the few galactic sites capable of sustaining the galactic cosmic
ray population against loss by escape, nuclear interactions and
ionization energy loss assuming a SN rate of about 1 per 30 years and
a 10\% efficiency for converting the mechanical energy into
relativistic particles.  Second, models of diffuse shock acceleration provide
a plausible mechanism for efficiently converting this explosion energy
into accelerated particles with energies $\sim 10^{14} - 10^{15}$~eV
and naturally give a power-law spectrum similar to that inferred from
the cosmic ray data after correcting for energy dependent propagation
effects.  Finally, observations of non-thermal X-ray emission in SN1006
(Koyama, et al., 1995) and IC443 (Keohane, et al., 1997) 
suggest the presence of electrons accelerated to $\sim 100$ TeV and
$\sim 10$ TeV respectively.

If SNRs are sites for cosmic ray production, there will be interactions
between the accelerated particles and the local swept-up interstellar
matter. Drury, Aharonian and Volk (1994) (DAV) and Naito and Takahara (1994)
have calculated the expected gamma-ray flux from secondary pion
production using the model of diffusive shock acceleration. The expected
intensity (DAV) is
\begin{equation}
F( > E) = 9\times10^{-11} \left(\frac{E}{1 {\rm TeV}}\right)^{-1.1}
                          \left(\frac{\theta E_{SN}}{10^{51}{\rm erg}}\right)
                          \left(\frac{d}{1 {\rm kpc}}\right)^{-2}
                          \left(\frac{n}{1 {\rm cm}^{-3}}\right) 
                           {\rm cm}^{-2}{\rm s}^{-1}
\end{equation}
where $E$ is the photon energy, $\theta$ is the
efficiency for converting the supernova explosion energy, $E_{SN}$, into
accelerated particles, $d$ is the distance to the SNR and 
$n$ is the density of the local ISM.

\section{OBSERVATIONS}
We report on the results of observations of six nearby SNR (W44,
W51, gamma-Cygni, W63, Tycho and IC443) by the Whipple Observatory's
high energy gamma-ray telescope situated on Mount Hopkins in southern
Arizona. The telescope (Cawley et al., 1990) employs
a 10 m diameter optical reflector to focus \v{C}erenkov light from
air showers onto an array of 109 photmultipliers covering a 3 degree
field of view. By making use of distinctive differences in the
lateral distribution of gamma-ray induced showers and hadronic induced
showers, images can be selected as gamma-ray like based on their
angular spread. The determination of the incident direction of the
selected gamma-ray like events is accomplished by making use of
the orientation, elongation and asymmetry of the image. Monte Carlo
studies have shown that gamma-ray images are a) aligned towards their
source position on the sky b) elongated in proportion to their impact 
parameter on the ground and c) have a cometary shape with their light
distribution skewed towards their point of origin in the image plane.
Results on the Crab Nebula indicate that the angular resolution
function for the telescope using this technique is a
Gaussian with a standard deviation of 0.13 degrees.
(Lessard et al., 1997). A combination of Monte Carlo 
simulations and results on the Crab Nebula indicate that
the energy threshold of the technique is 300 GeV and the effective
collection area for a point source located at the center of the
field of view is $2.1 \times 10^8 {\rm cm}^2$ and is reduced for offset 
sources (Lessard et al., 1997).

\begin{wrapfigure}[31]{r}{6.5cm}
\epsfig{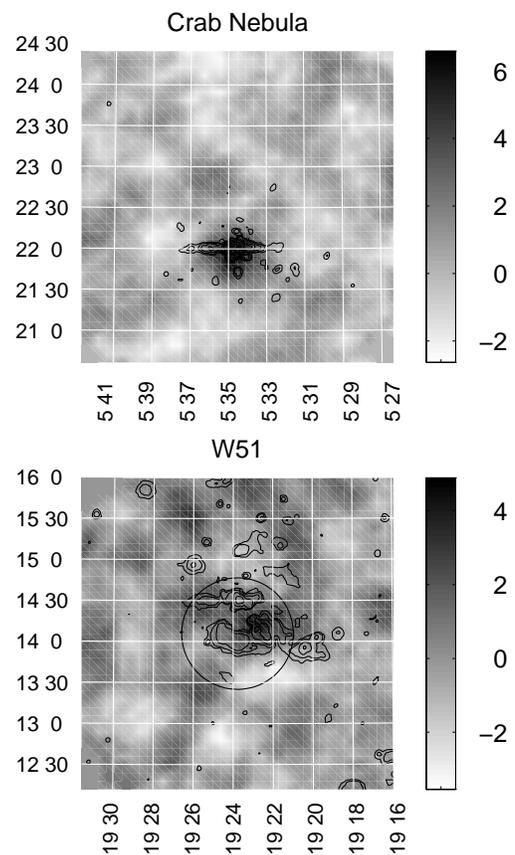}
\caption{Observations of the Crab Nebula and W51. We have applied a 
boxcar smoothing method which maximizes the point source sensitivity.
The statistical significance of the excess selected events is shown in 
grayscale. The overlayed contours are proportional to the 4850 MHz 
radio intensity.}
\end{wrapfigure}

The analysis of data from extended sources involves binning
the event arrival directions. We define the source region for
the SNR by a circular aperture which matches the maximum extent
of the radio shell (Green, 1995) plus twice the width of the angular resolution
function to account for the smearing of the edge of the remnant.
The number of gamma-ray candidate events is obtained by subtracting
the number of events in the OFF-source observations from the number
of events in the ON-source observations.

\section{RESULTS}
The observations were made over three observing seasons, from 1993 to
1996. Two dimensional images of the excess events for the Crab Nebula
(which was deliberately offset from the center of the camera, to
demonstrate the veracity of the technique) and the SNR W51 are shown
in Figure~1. In each frame, the statistical significance is displayed
in grayscale. The black contours are from the 4850 MHz radio survey by
Condon et al., 1994, showing the extent of the radio shell. The
circle shows the circular aperture used to derive the excess counts
from the entire remnant (see Table~1).

Of the six remnants observed, W 51 showed the greatest excess at an
offset location which is spatially consistent with hard x-ray emission
and peak radio intensity. More data are required to determine if this
excess is signicant. No significant excess has been recorded for the
other remnants and 99.9\% confidence upper limits on the flux have
been calculated (see Table 1). The upper limit 
assumes uniform emission from the
remnant in the absence of {\it a priori} knowledge of a more defined
emission region.

\begin{wrapfigure}[40]{r}{10cm}
\epsfig{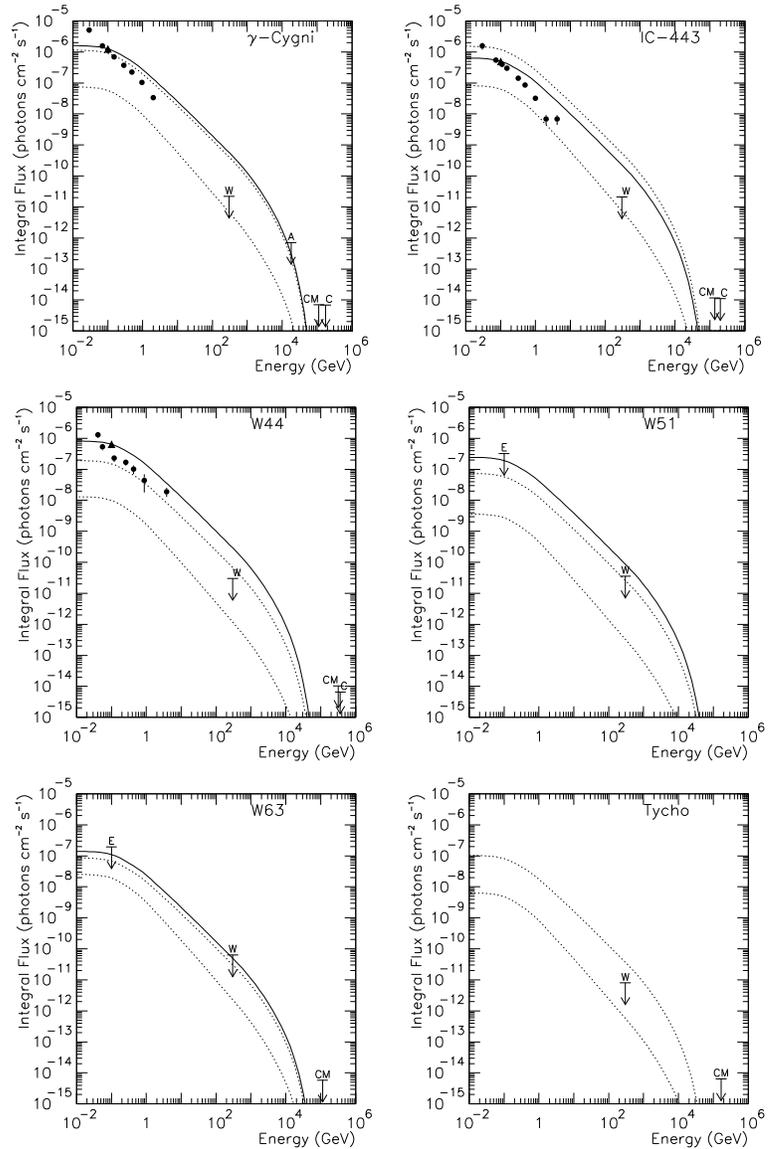}
\caption{Whipple upper limits shown along with EGRET integral fluxes, 
and integral spectra. These are compared to extrapolations from the
EGRET integral data points (solid curves), as well as a conservative
estimate of the allowable range of fluxes from the model of DAV 
(dotted curves). Also shown are CASA-MIA upper limits from Borione
et al., 1995, CYGNUS upper limits from Allen et al., 1995, 
and the AIROBIC upper limit from Prosch et al., 1996.}
\end{wrapfigure}

\section{DISCUSSION}
In Figure~2 the Whipple upper limits and EGRET data (Esposito
et al., 1996, for gamma-Cygni and IC443, Fierro, 1995, for W44,
and Thompson et al., 1995 for the remaining) are compared with
an $E^{-2.1}$ extrapolation of the EGRET data using the contribution
to the gamma-ray spectrum from secondary pion decay as derived by
Buckley et al. (1997) using the model of DAV. The upper dotted
curve assumes a source spectrum of $E^{-2.1}$ and a reasonable maximum
value of the product $E_{SN} \theta / d^2$ used
in the model. The lower dotted curve assumes a source spectrum of
$E^{-2.3}$ and a reasonable minimum value of the product
$E_{SN} \theta / d^2$.

We interpret our results in the context of two hypotheses, (1) that the
EGRET data gives evidence for acceleration of cosmic ray nuclei in SNR
and that the observed gamma-ray emission comes not from primary
electrons but from nuclear interactions of cosmic rays with ambient
material or (2) that the EGRET flux is produced by some other mechanism.

Under the assumption that the contribution from electron
brems\-stra\-hlung and inverse Compton (IC) scattering are neglible, it is
reasonable to compare the high energy gamma-ray upper limits to an
extrapolation of the integral EGRET fluxes using the model by DAV. In
the case of gamma-Cygni, IC443 and W44 the Whipple upper limits lie a
factor of $\sim 25$, 10 and 10 respectively below the extrapolation
and require either a spectral break or a source spectrum steeper than
$E^{-2.5}$ for gamma-Cygni and $E^{-2.4}$ for IC443.

Another plausible explanation for the results is that the EGRET flux
is produced by high energy electrons accelerated in the vicinity of
pulsars. If this is the case, then the Whipple upper limits must be
compared with the {\it a priori} model predictions. There is enough
uncertainty in the parameters of the SNR that the upper limits are not
in strong conflict with these predictions, but it is still strange
that in these objects which show strong evidence for interactions with 
molecular clouds (corresponding to the upper dotted curve) in no case
is there an observable TeV gamma-ray flux. Evidence of an 
X-ray point source embedded in gamma-Cygni
(Brazier et al., 1996) and IC443 (Keohane et al., 1997)
and the observation of a pulsar, B1853+01, in W44 (Wolszczan, et al.
1991), all provide support to a pulsar origin for the EGRET flux.

\begin{table}
\caption{Results of Observations.}
\vspace{0.5cm}
\begin{tabular}{lllllll} \hline \hline
Source & Pointing & Aperture & ON     & OFF    & Total & Upper \\
Name   & $\alpha,
         \delta$  & Radius   & Source & Source & Time  & Limit
                                                         $\times
                                                         10^{-11}$\\
       & (1950)   & (deg)    & Counts & Counts & (min) & $({\rm cm}^{-2}
                                                         {\rm s}^{-1})$\\
W44    & 18:53:29, 01:14:57 & 0.55     & 450    & 426    & 360.1 & 3.0 \\
W51    & 19:21:30, 14:00:00 & 0.68     & 619    & 559    & 468.0 & 3.6 \\
$\gamma$-Cygni& 20:18:59, 40:15:17 & 0.76     & 1040   & 1104   & 560.0 & 2.2 \\
W63    & 20:17:15, 45:24:36 & 1.05     & 452    & 501    & 140.0 & 6.4 \\
Tycho  & 00:22:28, 63:52:11 & 0.29     & 315    & 302    & 867.2 & 0.8 \\
IC443  & 06:14:00, 22:30:00 & 0.64     & 1565   & 1522   & 1076.7& 2.1 \\
\hline \hline
\end{tabular}
\end{table}
\newpage
\section{ACKNOWLEDGEMENTS}
	We acknowledge the technical assistance of Kevin Harris and
Teresa Lappin.
This research has been supported in part in the U.S. by the Department of
Energy and NASA, Forbairt in Ireland and PPARC in the UK.

\section{REFERENCES}
\setlength{\parindent}{-5mm}
\begin{list}{}{\topsep 0pt \partopsep 0pt \itemsep 0pt \leftmargin 5mm
\parsep 0pt \itemindent -5mm}
\vspace{-15pt}
\item Allen, G.E., et al., 1995, ApJ, 448, L25.
\item Borione, A., et al., 1995, in Proc. 24th Int. Cosmic Ray Conf. (Rome), 2,439.
\item Brazier, K.T., et al., 1996, MNRAS, 281, 1033.
\item Buckley, J.H., et al., 1997, in preparation.
\item Cawley, M.F., et al., 1990, Exper. Astr., 1, 173.
\item Condon, J.J., et al., 1994, AJ, 107, 1829.
\item Drury, L.O'C., et al., 1994, AA, 287, 959.
\item Esposito, J.A., et al., 1996, ApJ, 461, 820.
\item Fierro, J.M., 1995, PhD Thesis.
\item Green, D.A., 1995, A Catalog of Galactic Supernova Remnants (1995 July
      version), Cambridge, UK, Mullard Radio Astronomy Observatory, Available
      on the World Wide Web at \\
      http://www.phy.cam.uk.ac/www/research/ra.SNRs/snrs.intro.html.
\item Keohane, J.W., et al., 1997, accepted by ApJ.
\item Koyama, K., et al., 1995, Nature, 378, 255.
\item Lessard, R.W., 1997, Ph.D thesis, National University of Ireland.
\item Naito, T. and Takahara, F., 1994, J.Phys. G:Nucl.Part.Phys.,20,477.
\item Prosch, C., et al., 1996, AA, 314, 275.
\item Thompson, D.J., et al., 1995, ApJS, 101, 259.
\item Wolszczan, A., et al., 1991, ApJ, 372, L99.
\end{list}

\end{document}